\documentstyle[aps,psfig,times]{revtex}
\begin{document}
\title{ 
Synchronous clusters in a noisy inhibitory neural network}
\author{P.H.E. Tiesinga$^{1,2}$\footnote{Corresponding author, Tel: 619 453 4100 ext 1039,
Fax: 619 455 7933, E-mail: tiesinga@salk.edu} 
and Jorge V. Jos\'e $^{1}$\\
$^1$ Center for Interdisciplinary Research on Complex Systems,\\
and Department of Physics, Northeastern University, Boston
MA 02115, USA.\\
$^2$ Sloan Center for Theoretical Neurobiology, Salk Institute,\\
 10010 N. Torrey Pines Rd.,
La Jolla, CA 92037.}
\maketitle
\newpage

\newpage
\begin{abstract}
We study the stability and information encoding capacity
of synchronized states in a neuronal network model that represents
part of thalamic circuitry. Our model neurons have
a Hodgkin-Huxley-type low threshold Calcium channel, display
post inhibitory rebound, and are connected via GABAergic inhibitory synapses.

We find that there is a threshold in synaptic strength, $\tau_c$, below
which there are no stable spiking network states. Above
threshold the stable spiking state is a cluster state, where
different groups of neurons fire consecutively,
and each neuron fires with the same cluster each time.
Weak noise destabilizes this state, but
stronger noise drives the system into a different, self-organized,
stochastically synchronized state. Neuronal firing is still organized
in clusters, but individual neurons can hop from cluster to cluster.
Noise can actually induce and
sustain such a state below the threshold of synaptic strength.
We do find a qualitative difference in the firing patterns between small ($\sim 10$ neurons)
and large ($\sim 1000$ neurons) networks.

We determine the information content of the spike trains
in terms of two separate contributions: the spike time jitter around cluster firing times,
and the hopping from cluster to cluster. We quantify the information loss
due to temporally correlated interspike intervals.
Recent experiments 
on the locust olfactory system and striatal 
neurons suggest that the nervous system may actually use these  two channels to encode separate and unique information. 
\end{abstract}
\newpage
\twocolumn
\section{Introduction}
The brain receives an enormous amount of information transduced
by peripheral sense organs. This massive information influx is
coded and decoded in ways that are not yet fully understood in
cognitive neuroscience. Recent studies  have focussed on specific neural
substrates for binding mechanisms. Binding is the process by which
the brain combines different aspects of sensory modalities of one
object into one unified percept. 
The neural mechanisms that underlie synchronization
in different parts of the brain are only
partly understood. There is the suggestion that 
synchronization  may be relevant to binding \cite{binding3}.
Recent experiments have shown that inhibitory interneurons
in the hippocampus \cite{Whittington}, the thalamic reticular nucleus
\cite{Steriade93}, and the locust olfactory system \cite{Laurent96a}
can indeed synchronize neuronal discharges.
Subsequent theoretical analysis of
networks of interneurons has shown that
strong synchronization by mutual inhibition
is only moderately robust against
neuronal heterogeneities \cite{wang} and
synaptic noise \cite{CNS}.
In strong synchronization  all the neurons fire with a
short time-interval from each other.

In most experiments to date one measures the activity
of one neuron, or a small population of neurons.
Periodic oscillations (extracellular, or subthreshold intracellular)
measured in these experiments are
consistent with strong as well as weak synchronization.
In weak synchronization the {\em average} neuronal activity is
periodic, without each individual neuron
having to fire at each period. Often theoretical analyses, however,
have focussed on strong synchronization.
Here we conjecture
that weak synchronization is robust against neuronal heterogeneities
and synaptic noise, and consequently it is much more likely to occur in
neuronal systems. Furthermore we show that it can encode more information
compared to strongly synchronized states.
We present numerical results of weak
synchronization in a simple model of a network of Thalamic neurons
that supports our conjecture. We use a thalamic network, as an example,
due to the wealth of modeling information that is already available.
The mechanism we discuss here, however, has more general applicability.

The thalamus acts as a relay for most of the sensory information
that travels to cortical structures.
It regulates sleep-wake cycles \cite{Steriade93} and it may be involved
in early stimulus binding \cite{Sillito94}.
The lateral geniculate nucleus (LGN) and
thalamic reticular nucleus (TRN)  that are involved in vision have been 
studied extensively.
Neurons of the thalamus express low threshold Calcium currents 
\cite{Llinas84b}, and
they rebound after a sustained hyperpolarization.
It has been shown experimentally and in model calculations
that inhibitory neurons can synchronize neuronal discharges
in the thalamus
\cite{wangrinzel93,wangetal95,GolombRinzel94,Krosigk93a,Bal95a} and
produce traveling waves \cite{Rinzel98,Destexhe96,Destexhe96c,GolombRinzel96}.

The hallmark of weak synchronization is multimodal interspike interval
(ISI) histograms (ISIH). The ISI occurs only near multiples
of a particular time-scale, e.g. the period of the population
activity $T$. Multimodality of the ISIH has
 been observed
in the LGN \cite{Funke96}, and it was attributed to the action
of inhibitory neurons.
Multimodal ISIH have also been found in model simulations
of coupled inhibitory networks in the presence of noise \cite{Golomb92}
and in systems exhibiting stochastic 
resonance (SR) due to a periodic drive \cite{Wies95}.
A theoretical mechanism for autonomous
stochastic resonance (ASR) was proposed in a recent paper \cite{Longtin97}.
There the periodic drive was replaced by a periodic
mode in an internal kinetic variable, such that
the spikes ride on top of subthreshold voltage
oscillations. In our work the periodic neuronal activity
in the noise-driven system is self-induced by the network.
This mechanism is absent in unconnected single neurons,
or in a single element with autosynaptic feedback.

It has been suggested that the brain may encode
information through an ensemble or cluster of neurons
that fire within a short time of each other \cite{Laurent97,Riehle97}.
A particular neuron
may be part of a cluster for a few cycles, before it joins another
neuronal ensemble.  This type of dynamics is very similar to the
neuronal clusters that form in our model simulations described below.
An important problem is how to quantify the information
content of these binding-like cluster states. The Shannon entropy has
been used as a measure of information content in investigations
of sensory neurons in, for instance, crickets \cite{Levin96},
and flies \cite{Rieke97}. It is, nonetheless, not known
how the brain processes information, and thus it is not clear
whether the Shannon entropy is the correct quantity for this purpose.
It does, however,
provide an upper bound on the theoretical information content of spiking neurons.
It also implies that noise in the nervous system contains information, and
that noisy neurons are transmitting more information compared to regular
noiseless spiking neurons.
We emphasize that this statement is still controversial 
\cite{Softky93,Shadlen94,Shadlen98},
because even if the entropy measure yields consistent results in 
sensory systems this
does not guarantee its relevance to the central nervous system.
With these caveats in mind we still proceed to characterize the information
content of our neural networks by calculating its well-defined 
Shannon entropy.

One would like to calculate both the output entropy of the model system and 
the mutual information. The mutual information quantifies how the
ensemble of outputs is related to one of the possible realizations of the
input, and it involves additional averagings over a conditional
probability distribution which makes it very hard to calculate.
Even the simpler calculation of the Shannon entropy from its 
definition in terms
of the spike times is a difficult calculation.
Here we shall focus on the Shannon entropy of the neuronal output of a neuron
as part of the complete network.
We also present here some approximations that allows us to estimate the Shannon
entropy using the interspike interval time series.

\section{Methods}
\subsection{Network model}
Our single neuron model equation contains a low threshold Calcium current $I_{Ca}$,
a general leak current $I_L$, a synaptic current $I_{syn}$, and 
a noise current $C_m\xi$, 
\begin{equation}
C_m \frac{dV}{dt}=-I_{Ca}-I_L-I_{syn}-C_m\xi,
\label{VOLTEQ}
\end{equation}
together with the first order (Hodgkin-Huxley type) kinetic equations 
for the activation $m$ and inactivation $h$ variables for $I_{Ca}$
and  the synaptic variable $s$. This yields a neuronal dynamics
in terms of four variables, $V$, $m$, $h$, and $s$.
We have used the kinetics for $I_{Ca}$ and $I_{syn}$ as specified 
in \cite{Rinzel98} (a detailed description of the model is given in Appendix A). 
Our single neuron model captures some important features of the dynamics
of thalamic neurons, in particular its post inhibitory rebound (PIR).
We are presently studying a more complete model,
incorporating thalamo-cortical relay neurons
and GABAergic thalamic reticular neurons \cite{Destexhe96}, including all the
relevant active currents \cite{Huguenard92,McCormick92}. Our preliminary
results suggest that this does not change the conclusions
of our discussion here.

The neurons in our network are connected all-to-all by inhibitory GABAergic
synapses. Previous studies \cite{Destexhe96c,GolombRinzel96}
have shown that the precise spatial connectivity is important for 
the activity propagation.
In this work,  we will not consider the spatial characteristics
of the neuronal activity.
We have studied different sized systems, varying from $N=1$ (a single
neuron with autosynaptic feedback) to $N=1000$.
We also have included two types of noises in our model,
either Gaussian current noise, characterized by $\langle \xi \rangle=0$ and
\begin{equation}
\langle \xi(t) \xi(0) \rangle=2D\delta(t),
\end{equation}
with D the strength of the noise, 
or with Poisson distributed excitatory post-synaptic potentials (EPSPs)
and inhibitory post-synaptic potentials (IPSPs).
In our previous work we have shown that these two types of noises are not 
fully equivalent \cite{CNS98}. Both can generate, however, similar statistics,
and theoretically Gaussian noise is easier to control and vary.
The results presented here are thus obtained with Gaussian noise.
Unless stated differently the physiological total synaptic conductance used is
$g_s=2~\mbox{mS/cm}^2$, and the decay time 
of the synaptic channel $\tau_s=16~\mbox{ms}$.
The noise strength D is expressed in units of $mV^2/ms$,
time in ms, currents in $\mu A/cm^2$, and voltage in mV.

The resulting differential equations for $V_i$, $m_i$, $h_i$, $s_i$ are
numerically integrated
using a noise-adapted second order Runge-Kutta algorithm \cite{CNS} with
a time-step $dt=0.1~\mbox{ms}$.
The calculation starts with random initial conditions, with
the initial voltage chosen from a uniform distribution with a range 
of $20~\mbox{mV}$ centered around $-68~\mbox{mV}$, and $m$, $h$, and $s$  
are set equal to
their asymptotic values for a given value of $V$.

\subsection{Calculated quantities}
The raw model output are the time-traces for $V_i$, $m_i$, $h_i$, and $s_i$.
The spike-times
are defined as the time when the voltage $V_i$ crosses $-30~\mbox{mV}$ 
from below. We determined the standard histograms of interspike intervals 
\cite{Gerstein62}. 
The instantaneous firing rate, or frequency f, is defined
as the number of action potentials per second in a bin of $2~\mbox{ms}$.
Both the ISI histogram and $f$ are averaged over all neurons in the network.
We also calculated
\begin{equation}
v_{syn}=\frac{1}{N}\sum_i s_i,
\label{VSYN}
\end{equation}
which is proportional to
the current drive due to the synaptic connections with other neurons (and
itself).
Because the network is connected all-to-all, $v_{syn}$ is the same for
each neuron and represents an averaged or mean field type drive.
The variable $h$ determines the excitability of the neuron,
and the state of the network
strongly depends on the $h$-value distribution.
When $v_{syn}$ is below a certain value the network is disinhibited
and will fire shortly. To determine the $h$-distribution prior
to firing we have chosen $v_{syn}=0.01$ as the threshold.
This value is reached every cycle, except in the presence of
strong noise. In that case the disperse nature of the firing 
creates an average value of $v_{syn}$ above $0.01$.
We have determined both the instantaneous as well as the time-averaged
distribution of $h$.

\section{Results}

\begin{figure}
\unitlength=0.1in
\begin{picture}(40,80)
\includegraphics{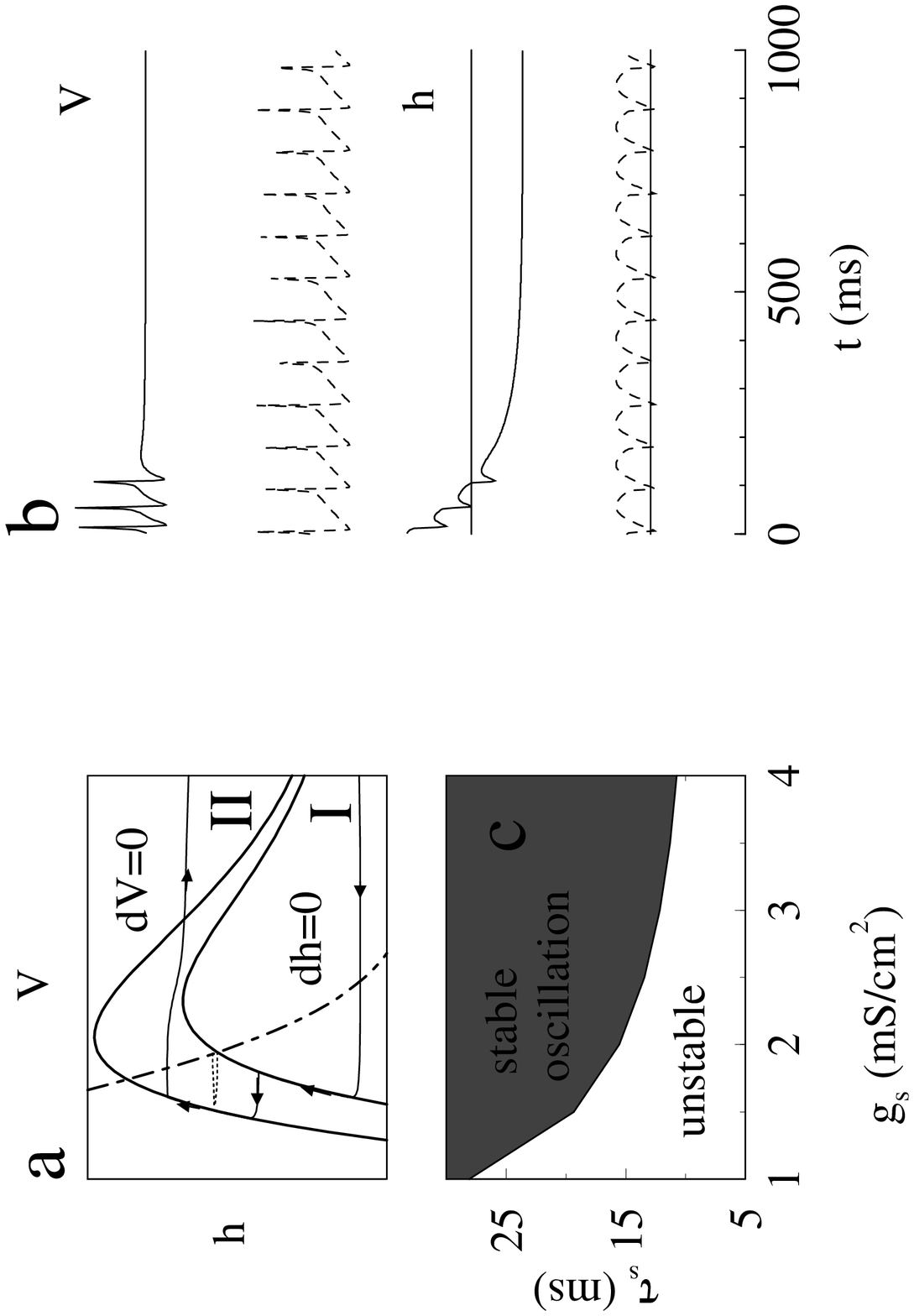}
\end{picture}
\caption{
(a) Phase-plane plots of the neuronal dynamics for the variables $h$ and $V$.
The plotted nullclines correspond to $dh/dt=0$ (dot-dashed line), and $dV/dt=0$ (solid lines), 
$I=-1.0$ (top) and $I=0.0$ (bottom). The phase trajectory of a
neuron released 
after $10~\mbox{ms}$ (dotted),
and  $200~\mbox{ms}$ (dashed) with hyperpolarizing current pulse $I=-1$. 
(b) Top curves for $V$, and bottom curves for $h$, plotted as a function of time
for $\tau_s=5~\mbox{ms}~<\tau_c$(solid lines), and curves $\tau_s=16~\mbox{ms}~>\tau_c$  (dashed), with $g_s=2.0$.
(c) Phase-diagram $\tau_s$ vs. $g_s$. Stable oscillations for
$\tau_s>\tau_c(g_s)$ are shaded while the unstable ones $\tau_s<\tau_c(g_s)$
are colored white.
}
\label{Fig1}
\end{figure}

The model considered here contains an inward low threshold Calcium current, $I_T$,
that initiates the Calcium action potentials. It is
inactivated at the resting membrane potential (RMP, equal to $-65.57~\mbox{mV}$), and
it is de-inactivated at hyperpolarized voltages. For the neuron to be
excitable, $h$ has to be de-inactivated (i.e. $h>0.305$).
We illustrate this in Fig.\ref{Fig1}(a). There are
two V null-clines drawn, one (I) in the absence of a current, and the other
(II) in the presence of a constant hyperpolarizing current $i=-1~\mu A/cm^2$. We apply
a short (10ms) and a long pulse (200ms) with strength $i$. The phase
point moves to II, and the $h$ value starts increasing with time-scale
$\tau_1=500~\mbox{ms}$ (see Appendix A). Upon termination of the short pulse the 
phase point moves back to I, without generating an action potential (AP). When the
long pulse ends, however, the $h$-value is too high, and the phase point misses
the nearby branch of I, and generates an AP.

\begin{figure}
\unitlength=0.1in
\begin{picture}(40,80)
\includegraphics{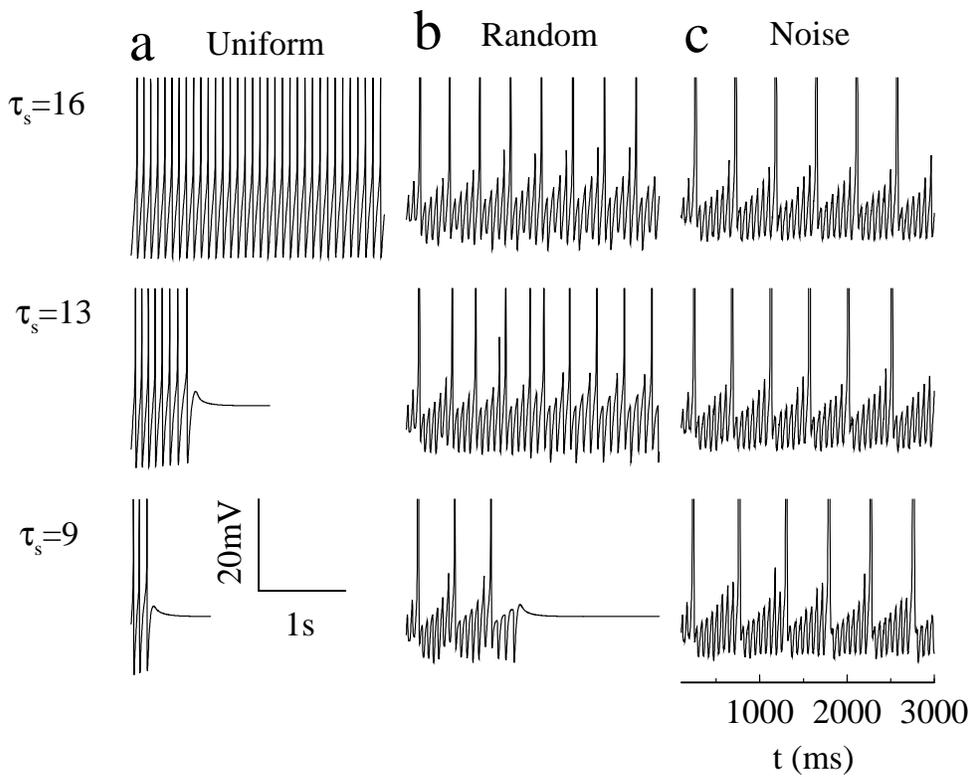}
\end{picture}
\caption{
Voltage traces for neuron one of an $N=1000$ neuron network of all-to-all
connected with (a) uniform initial conditions,
(b) random initial conditions, (c) uniform initial conditions
with noise ($D=0.02$, without the transient of $100~\mbox{ms}$),
for three different values of $\tau_s=16$, $13$, $9$ (from top to bottom).
A voltage and time scale bar is shown in the lower left graph.
}
\label{Fig2}
\end{figure}

The  necessary hyperpolarization is supplied by the inhibitory
postsynaptic potential (IPSP) generated by the activity of
other neurons in the network.
The strength of the IPSPs is determined by the
value of the synaptic conductance $g_s$ and the decay time $\tau_s$.
The critical value for 
periodic oscillations is defined as $\tau_s=\tau_c(g_s,N)$, that depends on
the number $N$ of neurons in the network.
For $\tau_s<\tau_c$ the oscillation dies out after a finite number
of action potentials. 
In Fig.$~$\ref{Fig1}(b) we show the voltage trace oscillations below
and above threshold for $N=1$ (single neuron
with autosynaptic feedback). At the start of the simulation
 the neuron is released from a hyperpolarized voltage.
 For subthreshold values of $\tau_s$ the
neuron produces a few spikes before returning to RMP. During
each spike the average $h$ value decreases, since the
inhibitory drive is not strong enough to replenish the loss due
to the depolarization of the AP.
Above threshold a periodic spike train is produced.
The $h$-value varies periodically, it decreases during
the AP and it increases during the inhibition.
Note that the interspike intervals are determined by $\tau_s$.
The subthreshold spike train has therefore a much smaller ISI compared
to the one above threshold.
We have determined the boundary between stable and unstable
oscillations as function of $g_s$ (Fig.$~$\ref{Fig1}(a)). 
The minimum
duration $\tau_s$ of the IPSP needed for deinactivation, increases
for weaker synaptic coupling $g_s$.
The situation for a real network ($N>1$) is more complicated, since 
the initial voltages play an important role.
If, for instance, we would start with neurons clamped at their resting membrane
potential nothing will happen.
To obtain a spiking network state we therefore always start the simulations
with part or all the neurons clamped at
hyperpolarizing voltages. 
With uniform initial conditions
all neurons are clamped at the same voltage value.
The threshold $\tau_c$ for self-sustained oscillations is then equal
to the one for a single neuron (Fig.\ref{Fig2}(a)).
 Above threshold this network
is in a coherent state: all neurons spike at the same time.
For random initial conditions the initial voltage is chosen
from a uniform distribution with a range of $20~\mbox{mV}$ around
a hyperpolarized average value.
In that case the network can sustain stable
oscillations for lower values of $\tau_s$ (Fig.\ref{Fig2}(b)). 
The network settles in a state where groups of neurons
fire simultaneously. It is easy to understand why such
cluster states emerge. Starting from random initial conditions
each neuron will have a different phase, and will thus 
reach the AP threshold at a different time. The first neurons to fire
will cause an inhibition that blocks other neurons (further from
threshold) from firing.
They can only fire after the decay of the inhibition (a few $\tau_s$).
Periodic oscillations in the network are sustained by
inhibition waves produced by the activity of de-inactivated
neurons. The oscillations automatically
become coherent, with the initial phase differences between 
cluster neurons driven to zero.
In the simplest cluster state  each neuron
fires with the same ISI. The time between consecutive
cluster firings, or cycle length,  may vary since the strength of
inhibition depends on the cluster
size. Neurons will only fire when $v_{syn}$ is below a certain value:
the higher the initial value (proportional to cluster size),  the
longer it takes to reach this value. More complex cluster states may also
contain neurons that fire with different frequencies.

\begin{figure}
\unitlength=0.1in
\begin{picture}(40,80)
\includegraphics{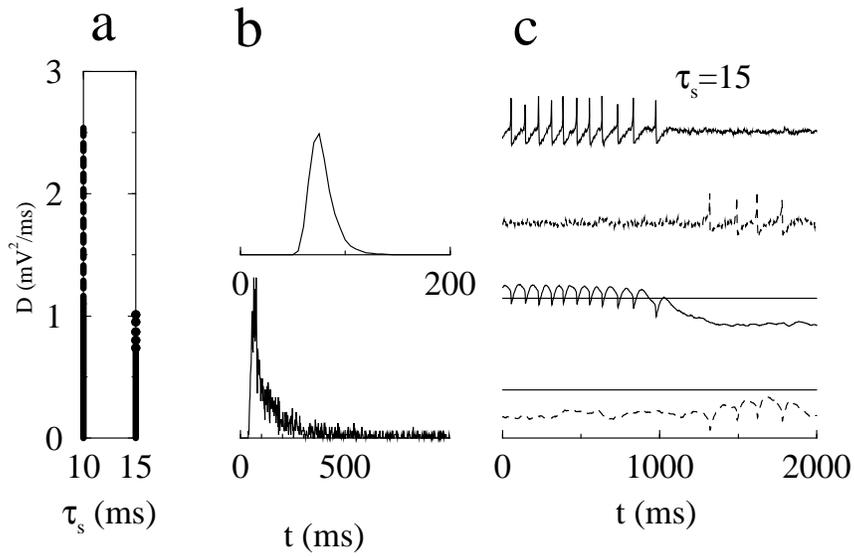}
\end{picture}
\caption{
Noise-driven single neuron dynamics with autosynaptic feedback.
(a) Phase-diagram of neuronal behavior as function of noise strength $D$,
for different values of $\tau_s$.  
No activity (solid line), spontaneous single spikes (dashed line),
spontaneous spike-trains (filled circles).
(b) ISIH for (top) $\tau_s=15$  and $D=0.76$;
(bottom) $\tau_s=10$ and $D=2$.
(c) Representative voltage (top) and $h$ (bottom)  time traces
for $D=0.26$ (solid lines) and $D=0.76$ (dashed lines) with $\tau_s=15$.
}
\label{Fig3}
\end{figure}

Sufficiently strong noise
can induce and maintain a spiking network state even for $\tau_s<\tau_c$.
Again we have to distinguish between single neurons and a network.
The noise-induced dynamics of a single neuron 
with autosynaptic feedback will not yield a periodic spike-trace.
Instead, the ISI distribution  
has a peak for short times due to ISIs within the spike trains,
and an exponential distribution for the intervals between the end
of one, and the start of another spike train.
We show some representative voltage  traces in Fig.$~$\ref{Fig3}(c).
Close to threshold and with weak noise the neuron
produces a long transient that dies
out eventually. Stronger noise can spontaneously induce a spike.
The inhibition induced by the AP then manages to produce a short spike train.
The number of spikes in this train depends on the distance from
threshold. 
This makes the dynamics discussed here essentially different from Stochastic Resonance, since
in that case one would
obtain a multimodal distribution, with peaks at multiples of the driving frequency.

\begin{figure}
\unitlength=0.1in
\begin{picture}(40,80)
\includegraphics{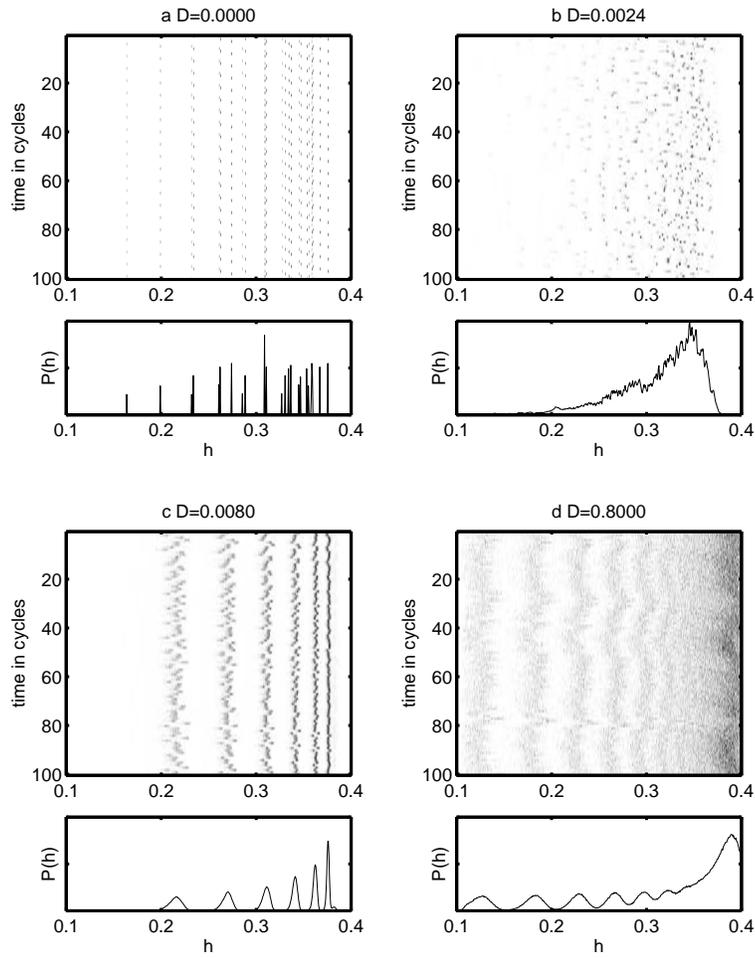}
\end{picture}
\caption{
Each panel consists of a gray-scale coded 
instantaneous $h$ value distribution for consecutive cycles (top),
and the time-averaged $h$ distribution (over at least 500 cycles) (bottom).
From left to right, top to bottom the noise-values are 
$D=0$, $0.024$, $0.08$, $0.8$. In the upper left panel (a) 
the delta function peaks in the h-distribution have been broadened to
enhance visibility. 
}
\label{Fig4}
\end{figure}

\begin{figure}
\unitlength=0.1in
\begin{picture}(40,80)
\includegraphics{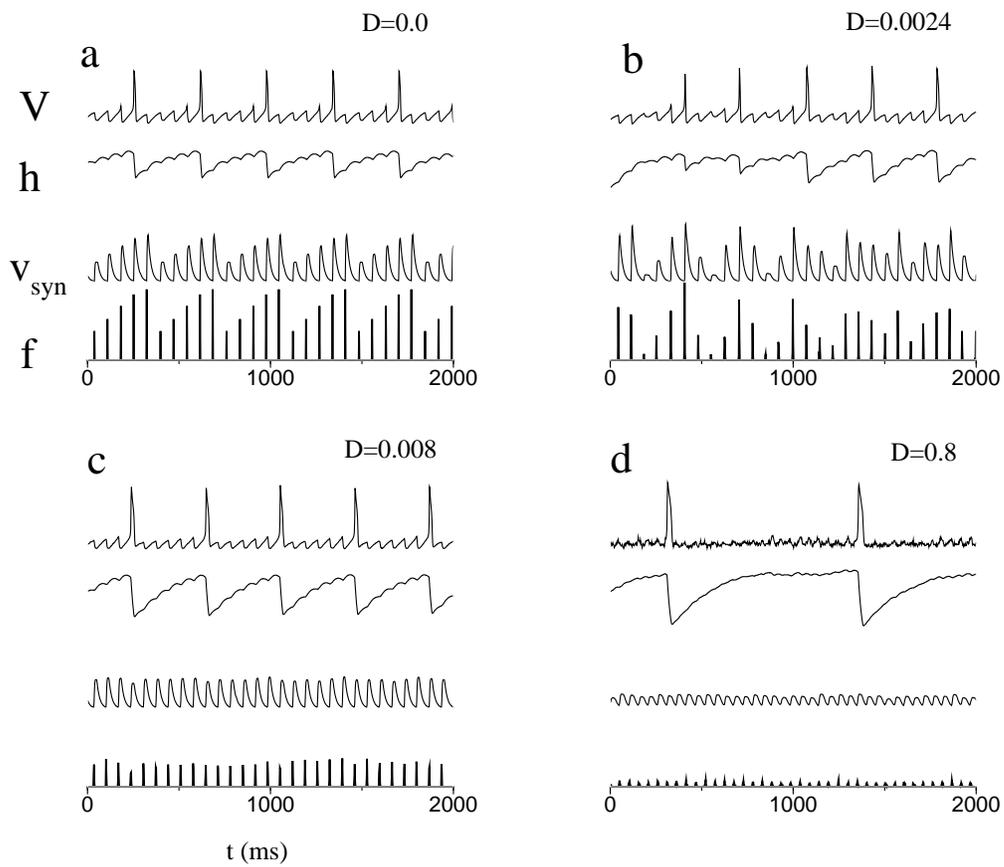}
\end{picture}
\caption{
Four panels with the same $D$ values given in Fig.$~$\ref{Fig4}:
from left to right, top to bottom the noise-values are
$D=0$, $0.024$, $0.08$, $0.8$.
We plot in each panel from top to bottom 
the $V$ and $h$ functions of neuron one, the population average $v_{syn}$
of the synaptic variables $s_i$, and the instantaneous firing rate $f$ as 
a function of time.
}
\label{Fig5}
\end{figure}

We studied the dynamics of the cluster states in a network with $N=1000$,
in terms of the variables shown in Figs.$~$\ref{Fig4} and $~$\ref{Fig5}.
We focused on four values for $D=0$, $0.0024$, $0.008$, and $0.8$.
An important variable in our analysis is $h$, the inactivation variable of
$I_T$, since the model neuron is only excitable when $h$ is large
enough. The distribution of $h$ values in the network
will tell us which neurons are excitable, and which ones
need to be de-inactivated by further inhibition cycles.
The regularity of the network dynamics is further reflected in the periodicity 
of $f$ and $v_{syn}$ (Fig.$~$\ref{Fig5}) 
and their autocorrelation function (not shown).
Note that $v_{syn}$ itself acts as a drive on the neurons. The
larger the distance between the peak and trough, the stronger
the synchronizing force.

We can identify four different types of regimes.
For zero noise (D=0) the system is in a state with five
clusters of unequal size.
The distribution of cluster sizes is determined by the initial conditions.
At each time a neuron can
only have one of five h-values. This set of h-values
goes through a modulation with a period of five cycles (Fig$~$\ref{Fig4}(a)).
The derived quantities V, h, $v_{syn}$, and f (Fig.$~$\ref{Fig5}(a)) 
go through the same modulations.
When all the clusters have the same size, there are no such
modulations, and the $h$ histogram would consist of only $5$ peaks.

Immediately after firing, the neuron is partially de-inactivated 
with each subsequent cycle until it is excitable
again (see the $h$ time traces in Fig.$~$\ref{Fig5}).
The neuron then has to wait its turn to
become disinhibited and fire before other clusters do.
When there is noise, there is dispersion in the
spike firing times.
In the absence of time delay there is only a short time interval during
which neurons can fire before the inhibition blocks all other
firings during one cycle. As a result, clusters lose neurons
whose AP has been noise delayed, and other
clusters gain those neurons as members. In addition, noise can
cause neurons to fire before the rest in their cluster.

Weak noise (D=0.0024) disorders the system.
The network starts out with unequal cluster sizes (due to the initial
conditions).
Large clusters lose more neuron members than smaller
ones. Weak noise, however, is not strong enough to equalize
their numbers on the time-scale considered ($5\,\times10^4~\mbox{ms}$). Instead
the cluster sizes start to vary in a somewhat stochastic
fashion, leading to an erratic firing rate (Fig.$~$\ref{Fig4}(b))
and a fluctuating period. The time-averaged $h$ histogram is
very broad (Fig.$~$\ref{Fig5}(b)).

For stronger noise (D=0.008) the average cluster size
becomes stationary after a brief transient. The $h$-values that the neurons of different clusters cycle go through
are, on the average, the same. As a result there are six smooth peaks
in the $h$-histogram.
The peaks become sharper for more de-inactivated values,
and the h-distribution is stationary (on the average 
it is the same for each cycle). The actual neurons
that fire in each cluster changes with time.
This state is stable up to a noise strength of approximately D=1
(for N=1000).

The cycle-to-cycle fluctuations in cluster sizes increases with
increasing $D$.
The inhibition that
each cluster receives (proportional to $v_{syn}$) varies,
and as a result the width of the peaks in the $h$ histogram increases.
The cluster size also decreases with increasing $D$, and
the amplitude of $v_{syn}$ oscillations also decreases.
 The
stability of a cluster can be defined as the fraction of neurons
that are still part of a given cluster the next time it fires. This
is related to the number and height of the peaks in the ISIH
(see Fig.$~$\ref{Fig7}(a) and (b)), and it decreases with D. The neuron spends most
of its time in an excited state  (flat part of $h(t)$, Fig.$~$5(d))
waiting for its noise induced
AP threshold to fall in the inhibition free window. For that reason
it is unlikely to fire with the same cluster as in the previous time.
Finally, for larger noise strengths, $D>1$, the firing is no longer organized
in clusters, since 
the noise has become so large that during an inhibition free window 
not enough neurons fire coherently to create a large enough inhibition
to block the discharge until the next cycle. As a result there
are no quiescent periods defining cycles and no distinct cycles either.

\begin{figure}
\unitlength=0.1in
\begin{picture}(40,80)
\includegraphics{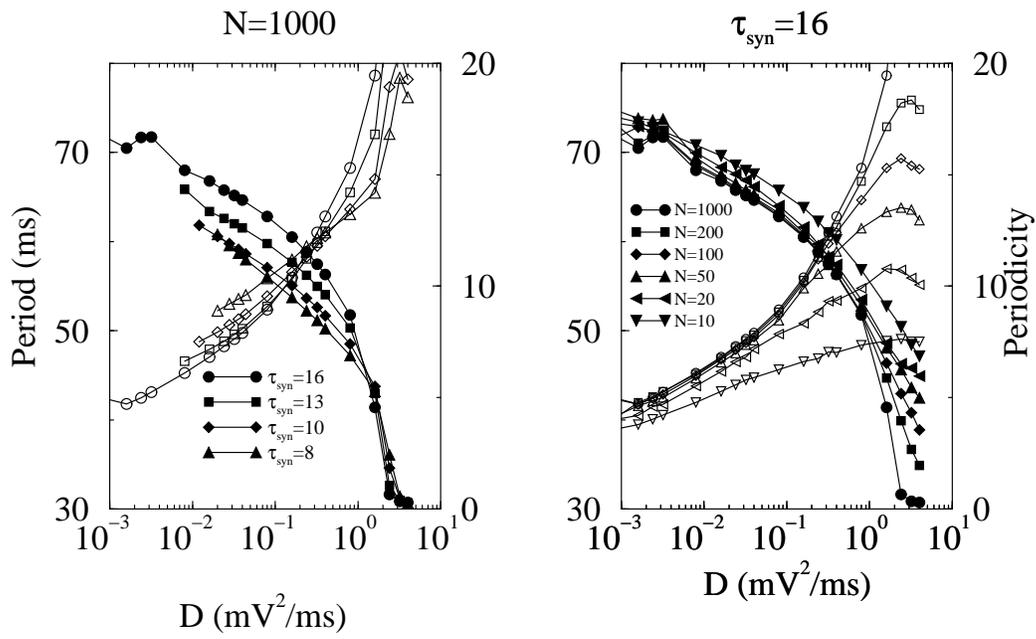}
\end{picture}
\caption{
Here we plot the average period (filled symbols, left hand scale),
 and the periodicity (open symbols, right hand scale)
as a function of $D$, for $\tau_s=8$, $10$, $13$, $16$
(for $N=1000$) and for different system sizes $N=10$, $20$, $50$, $100$,
$200$, and $1000$.
}
\label{Fig6}
\end{figure}

The cluster states can be quantitatively described by the average 
cycle length (period), and the periodicity (number of cycles) 
of the response.
We have studied these two quantities as a function of
D for different values of $\tau_s$ for N=1000,
and for different system sizes for $\tau_s=16$. 
For smaller $\tau_s$ the amount of inhibition available
for de-inactivation is lower, and the periodicity increases 
since the de-inactivation has to be spread out over more cycles.
The cycle-period, being proportional to $\tau_s$, also decreases.
In addition, the oscillation
needs a higher minimum value of $D$ to sustain itself, 
since the average cluster size is given by
the system size over the periodicity. When it becomes too small, the
periodic component in the inhibition 
becomes too small, and as a result the cluster state dies.
The noise strength for which this happens is only weakly dependent 
on $\tau_s$.
Note that the periodic component is proportional to the cluster size
normalized by the network size, in addition it
decreases with increasing jitter.
There is, however, a difference between large networks ($N\sim1000$)
and small ones ($N\sim 10$). For small networks, a single neuron
can provide enough inhibition to block neuronal discharge and thus 
to maintain a periodic network state. 
The state is very robust against noise, even when noise reduces 
the cluster to its minimum size (one neuron),
it can still maintain a spiking state. 
The drawback is that the fluctuations in cluster size will be of the
order of the cluster size itself, causing the cycle length to vary considerably.
Below threshold (not shown) the amount of noise needed to induce
a spiking state increases, and for very small networks ($N<4$) failure
can occur, i.e. the network becomes quiescent if one neuron fails to fire.
Note that the maximum periodicity that the system
can sustain is bounded by the system size. In Fig.$~$\ref{Fig6} we see that the periodicity
obtained for a given $D$ decreases with $N$.

We now discuss the possible information content of the ISI time series of 
individual neurons. We assume that the states are characterized by
a periodic population activity. This is reflected in the ISI time series
because the ISI will only take values close to multiples of the 
cycle length. The ISIH will therefore consist of a series of peaks.
If a neuron would consistently spike with the same cluster, there
would be only one peak. The peak with maximum weight is close
to the average ISI (and periodicity) of the network, the other peaks correspond
to the ISI where the neuron changed cluster. The relative weight of the maximum
peak is thus a measure of the stability of the clusters.
In Figs.$~$\ref{Fig7}(a), and \ref{Fig7}(b) we show 
the ISIH of a state with stable 
and unstable clusters, respectively.
The width of each peak represents the jitter around a multiple 
of $T$.
We find that the ISIH of an individual network neuron
is to a very good approximation the same as the population averaged ISIH.
This does not mean that all the neurons fire independently, it only states that
all individual neurons have identical properties, and that in the 
long run the statistics of their time-series are  the same.
Our analysis is therefore performed on the population averaged ISIH,
and we also use the population averaged return map.
We find that the ISIH is well described by a sum of Gaussians (SOG) of different
widths (see Appendix B), and that the width increases with the 
the peak number. In Appendix B we derive an expression given in 
Eq.$~$(\ref{SOGentropy}) for the
entropy of the SOG. We see that the number of peaks and their weight
--cluster hopping-- and their width --jitter within the cluster-- yield
two distinct contributions to the information encoding capacity.
In Fig.$~$\ref{Fig7}(c) we plot the entropy as a function of width $\sigma$ (taken constant for all peaks)
and the total number of $n$ peaks (weighted with a cosine envelope, 
Eq.$~$(\ref{CosEnv}). 
In our simulations we find that the amount of jitter and the number
of peaks are closely correlated, and grow with $D$.

Our present entropy calculations assume that there is no correlation
between consecutive ISIs. For a given amount of correlation $\gamma$
(see Appendix B) the entropy per spike will decrease. We have quantified
this in a model calculation for $n=2$ and various amounts
of correlation $\gamma$ (see Fig.$~$\ref{Fig7}(d)).

\section{Discussion}

In recent years considerable attention, as well as controversy, has been directed
at studying the variability of neuronal discharge in the cortex \cite{Shadlen98}.
It is beyond doubt that neurons {\em in vivo} are noisy. The question
is whether exact spike-times matter -- that is, if the jitter in spike times
represents information, for instance quantified by the
Shannon entropy -- or if only the average firing rate matters.
If spike times {\em do} matter, then the synchronized discharge
has a special significance.
An important question is whether
the nervous system is sensitive to synchronization or not.
Our work is relevant in shedding light to this fundamental question in two ways. We have shown
that noisy neurons can synchronize without the need of a strong
external drive, and that the synchronized neuronal
discharge has a potentially high information content. To place
our results in a proper context we will now discuss these points in more
detail below.

The role that inhibitory interneurons play in the functioning of the nervous system
has long been unclear. 
It is hard to find, and to record from interneurons, because they
are rather small. Moreover the output of the nervous system is mostly generated
by the principal neurons, so early investigations studied mainly the pyramidal
neurons. In recent years it has become clear
that inhibition plays a major role in synchronizing principal
neurons, in for example the hippocampus \cite{traub}, the thalamus \cite{Steriade93}, 
and the locust olfactory system \cite{Laurent96a}. 
The mechanism by which synchronized oscillations are generated in the brain is
only partly understood. {\em In vivo} many different rhythms of different
frequency have been observed. 
Pharmacological manipulations of slices in {\em in vitro} experiments
have elucidated some aspects of the synchronization mechanism.
For instance, Whittington {\em et al} \cite{traub} showed that 
GABA$_A$-mediated inhibition is responsible for the synchronization 
in hippocampal slices.
Slice experiments, however, suffer from the drawback that the
natural afferents are cut, and therefore  the synaptic activity
giving rise to the {\em in vivo} variability is absent.
Recent theoretical work has shown that in fact synchronization
by mutual inhibition is not robust against neuronal heterogeneities
\cite{wang} and synaptic noise \cite{CNS}.
In theoretical investigations one usually only considers strong synchronization.
In strong synchronization one imposes the strong constraint
that each neuron has to fire within a short interval from each other.
Here we propose that weak synchronization may in fact be more prevalent
in networks connected by chemical synapses. 
In weak synchronization the {\em average} neuronal activity is
periodic, without each individual neuron
having to fire at each period. 
Weak synchronization is for example consistent with the experiments
in \cite{traub}. There are exceptions. For example, recent experiments
on weakly electric fish show that neurons in the pacemaker nucleus are strongly
synchronized \cite{Moortgat98}. There, however, synchronization
can possibly be attributed to electric gap junctions.

Weak synchronization is associated with a periodic drive.
This drive is either generated externally, as is the case
with Stochastic Resonance \cite{Wies95,Gluck96},
or it is generated intrinsically by the network as it happens here.
The neuron then skips periods, which it can either do 
deterministically (usually one peak in ISIH),
or stochastically (multimodal ISIH).
In both cases the network dynamics consists of clusters of
neurons firing together. But in the latter case the neuronal composition
of the clusters varies with time.  A neuron that at a certain point
fired with a certain cluster A, 
can fire the next time with another cluster B.
This mechanism yields a synchronization that is robust
against noise, and neuronal heterogeneity.

Oscillating neural assemblies do play an important role
in the functioning of the invertebrate nervous system.
In a series of seminal experiments on the bee and locust olfactory system,
Laurent et al. have shown that different odors activate
overlapping ensembles of projection neurons \cite{Laurent94}.
The periodic discharge of the ensembles is coherent on a 
cycle-by-cycle basis. Odors may be classified from the
temporal firing pattern of projection neurons \cite{Laurent96b}.
Synchronization of the projection neurons may be abolished
by applying picrotoxin, and without changing their individual
response characteristics \cite{Laurent96a}. This desynchronization
was shown to impair the ability of bees to distinguish two closely related
odors \cite{Laurent97}, and subsequently a population of neurons was found
that was sensitive to the synchronization of projection neurons \cite{Laurent98}.
The ability to distinguish between two odors, 
based on their spike trains, was then reduced under desynchronized conditions.

We find that the information content, defined by the Shannon entropy of
the spike-time distribution, contains three contributions.
First, the jitter in the spike times around the cluster firing time.
Second, the distribution of the number of cycles between two consecutive
spikes, and finally the correlation between consecutive ISIs.
Our analysis has been performed on the (average) output of a single
neuron. One has to await the development of fast computational techniques
to tackle the more challenging problem of quantifying the information
output of the network, while taking into account the correlation in spike times
between different neurons due to the cluster state.
Our single-neuron spike-train results, however, may have direct relevance to recent
experimental work on striatal neurons \cite{Wilson98}. Striatal neurons
can be in the down (hyperpolarized, quiescent), or in the up-state
(depolarized, noisy). The transitions to the up-state are precisely
timed and synchronous. The fine-structure in the spike train
is asynchronous. Wilson {\em et al.} \cite{Wilson98} have suggested that
the brain may use these two channels to encode different types
of information.

We have studied the dependence of these oscillations on network
parameters. We find that there is difference between small (around ten neurons)
and large networks (a few hundred neurons), under the conditions of having a 
fixed total synaptic drive per neuron. 
For small networks one needs more noise to drive the subthreshold
network into stable oscillations. These oscillations are very 
robust against increases in the noise level, and the fluctuations in
the time between two cluster firings (cycle length) increases with the amount of noise.
For large networks strong noise causes an instability, the stable cluster size 
for a given amount of noise becomes too small to inhibit out of sync
neuronal discharges.  For intermediate noise-strengths the neuronal
dynamics self-organizes itself into a stochastically synchronized state.
We also find that the farther the network is below threshold, more
noise is necessary to induce a spiking state. 
The mechanism to create the oscillations is due to the competition
between the excitatory de-inactivating, and the inhibitory effect
of the synaptic drive. Each cycle will de-inactivate neurons, until
they are excitable again. The neuron then has to await the decay of
inhibition created by more excitable neurons. For some parameter
values the latter stage is absent, and the dynamics is fully deinactivation
dominated.
The important time-scales in the dynamics are the deinactivation
time-scale $\tau_1$ and the synaptic decay time $\tau_s$. The cycle or population
period scales directly with $\tau_s$. Our results therefore predict
that by pharmacologically decreasing $\tau_s$ one can increase the cycle frequency.

In summary, the brain has circuitry capable synchronizing
with heterogeneous components, and in the presence of noise.
The spike trains of the synchronized discharge still contain information.
Whether the brain utilizes this mechanism to synchronize, and more importantly
whether it uses the information in the precise temporal sequence 
is still an open question awaiting further study.
\section{Acknowledgements}
This work was partially funded by the Northeastern University CIRCS fund,
and the Sloan Center for Theoretical Neurobiology (PT). We thank TJ Sejnowski
for useful suggestions.


\newpage
\begin{thebibliography}{10}

\bibitem{binding3}
{Singer W} and {Gray CM}.
\newblock {Visual feature integration and the temporal correlation hypothesis}.
\newblock {\em Annu. Rev. Neurosci.}, pages 555--586, 1995.

\bibitem{Whittington}
{Whittington MA}, {Traub RD}, and {Jeffreys JGR}.
\newblock {Synchronized oscillations in interneuron networks driven by
  metabotropic glutamate receptor activation}.
\newblock {\em Nature}, 373:612--615, 1995.

\bibitem{Steriade93}
{Steriade M}, {Cormick DA}, and {Sejnowski TJ}.
\newblock {Thalamocortical oscillations in the sleeping and aroused brain}.
\newblock {\em Science}, 262:679--685, 1993.

\bibitem{Laurent96a}
{MacLeod K} and {Laurent G}.
\newblock {Distinct mechanisms for synchronization and temporal patterning of
  odor-encoding neural assemblies}.
\newblock {\em Science}, 274:976--979, 1996.

\bibitem{wang}
{Wang XJ} and {Buzs\'aki G}.
\newblock {Gamma oscillation by synaptic inhibition in a hippocampal
  interneuronal network model}.
\newblock {\em J. Neurosci.}, 16:6402--6413, 1996.

\bibitem{CNS}
{Tiesinga PHE}, {Rappel W-J}, and {Jos\' e, JV}.
\newblock { Synchronization in networks of noisy interneurons.}
\newblock In {Bower J}, editor, {\em {Computational Neuroscience}}, pages
  555--559. Plenum Press, New York, 1998.

\bibitem{Sillito94}
{Sillito AM}, {Jones HE}, {Gerstein GL}, and {West DC}.
\newblock {Feature-linked synchronization of thalamic relay cell firing by
  feedback from the visual cortex}.
\newblock {\em Nature}, 369:479--482, 1994.

\bibitem{Llinas84b}
{Jahnsen H} and {Llin\'as R}.
\newblock {Ionic basis for the electroresponsiveness and oscillatory properties
  of the guinea-pig thalamic neurones in vitro}.
\newblock {\em J. Physiol.}, 349:227--247, 1984.

\bibitem{wangrinzel93}
{Wang X-J} and {Rinzel J}.
\newblock { Spindle rhythmicity in the reticularis thalami nucleus:
  synchronization among mutually inhibitory neurons.}
\newblock {\em Neuroscience}, 53:899--904, 1993.

\bibitem{wangetal95}
{Wang X-J}, {Golomb D}, and {Rinzel J}.
\newblock {Emergent spindle oscillations and intermittent burst firing in a
  thalamic model: specific neuronal mechanisms.}
\newblock {\em Proc Natl Acad Sci USA}, 92:5577--5581, 1995.

\bibitem{GolombRinzel94}
{Golomb D} and {Rinzel J}.
\newblock {Synchronization properties of spindle oscillations in a thalamic
  reticular nucleus model}.
\newblock {\em J. Neurophys.}, 72:1109--1126, 1994.

\bibitem{Krosigk93a}
{von Krosigk M}, {Bal T}, and {McCormick DA}.
\newblock {Cellular mechanisms of a synchronized oscillation in the thalamus}.
\newblock {\em Science}, 261:361--364, 1993.

\bibitem{Bal95a}
{Bal T}, {von Krosigk M}, and {McCormick DA}.
\newblock {Synaptic mechanisms underlying synchronized oscillations in the
  ferret lateral geniculate nucleus in vitro}.
\newblock {\em J. Physiol.}, 483:641--663, 1995.

\bibitem{Rinzel98}
{Rinzel J}, {Terman D}, {Wang XJ}, and {Ermentrout B}.
\newblock {Propagating Activity Patterns in Large-Scale Inhibitory Neuronal
  Networks}.
\newblock {\em Science}, 279:1351--1355, 1998.

\bibitem{Destexhe96}
{Destexhe A} and {Sejnowski TJ}.
\newblock {Synchronized oscillations in thalamic networks: insights from
  modeling studies}.
\newblock In {Steriade M}, {Jones EG}, and {McCormick DA}, editors, {\em
  {Thalamus}}. Elsevier, 1996.

\bibitem{Destexhe96c}
{Destexhe A}, {Bal T}, {McCormick DA}, and {Sejnowski TJ}.
\newblock {Ionic mechanisms underlying synchronized oscillations and
  propagating waves in a model of ferret thalamic slices}.
\newblock {\em J. Neurophys.}, 76:2049--2070, 1996.

\bibitem{GolombRinzel96}
{Golomb D}, {Wang XJ}, and { Rinzel J}.
\newblock {Propagation of spindle waves in a thalamic slice model}.
\newblock {\em J. Neurophysiol.}, 75:750--769, 1996.

\bibitem{Funke96}
{Funke K}, {Nelle E}, {Li B}, and {W\"org\"otter F}.
\newblock {Corticofugal feedback improves the timing of retino-geniculate
  signal transmission}.
\newblock {\em Neuroreport}, 7:2130--2134, 1996.

\bibitem{Golomb92}
{Golomb D} and {Rinzel J}.
\newblock {Clustering in globally coupled inhibitory neurons}.
\newblock {\em Physica D}, 72:259--282, 1994.

\bibitem{Wies95}
{Wiesenfeld K} and {Moss F}.
\newblock {Stochastic resonance and the benefits of noise: from ice ages to
  crayfish and SQUIDs}.
\newblock {\em Nature}, 373:33--36, 1995.

\bibitem{Longtin97}
{Longtin A}.
\newblock {Autonomous stochastic resonance in bursting neurons}.
\newblock {\em Phys. Rev. E}, 55:868--876, 1997.

\bibitem{Laurent97}
{Stopfer M}, {Bhagavan S}, {Smith BH}, and {Laurent G}.
\newblock {Impaired odor discrimination on desynchronization of odor-encoding
  neural assemblies}.
\newblock {\em Nature}, 390:70--74, 1997.

\bibitem{Riehle97}
{Riehle A}, {Grun S}, {Diesmann M}, and {Aertsen A}.
\newblock {Spike synchronization and rate modulation differentially involved in
  motor cortical function}.
\newblock {\em Science}, 278:1950--1953, 1997.

\bibitem{Levin96}
{Levin JE} and {Miller JP}.
\newblock {Broeadband neural encoding in the cricket cercal sensory system
  enhanced by stochastic resonance}.
\newblock {\em Nature}, 380:165--168, 1996.

\bibitem{Rieke97}
{Rieke F}, {Warland D}, {{de Ruyter van Steveninck} RR}, and {Bialek W}.
\newblock {\em {Spikes: exploring the neural code}}.
\newblock MIT press, Cambridge, 1997.

\bibitem{Softky93}
{Softky WR} and {Koch C}.
\newblock {The highly irregular firing of cortical cells is inconsistent with
  temporal integration of random EPSPs}.
\newblock {\em J. Neurosci.}, 13:334--350, 1993.

\bibitem{Shadlen94}
{Shadlen MN} and {Newsome WT}.
\newblock {Noise, neural codes, and cortical organization}.
\newblock {\em Curr. Opin. Neurobiol.}, 4:569--579, 1994.

\bibitem{Shadlen98}
{Shadlen MN} and {Newsome WT}.
\newblock {The variable discharge of cortical neurons: implications for
  connectivity, computation, and information coding}.
\newblock {\em J. Neurosci.}, 18:3870--3896, 1998.

\bibitem{Huguenard92}
{Huguenard JR} and {McCormick DA}.
\newblock {Simulation of the currents involved in rhythmic oscillations in
  thalamic relay neurons}.
\newblock {\em J. Neurophys.}, 68:1373--1383, 1992.

\bibitem{McCormick92}
{McCormick DA} and {Huguenard JR}.
\newblock {A model of the electrophysiological properties of thalamocortical
  relay neurons}.
\newblock {\em J. Neurophys.}, 68:1384--1400, 1992.

\bibitem{CNS98}
{Tiesinga PHE} and {Jos\'e, JV}.
\newblock {Spiking Statistics in Noisy Hippocampal Interneurons.}
\newblock {\em Proceedings Computational Neuroscience 1998}, 1999.

\bibitem{Gerstein62}
{Rodieck RW}, {Kiang NY-S}, and {Gerstein GL}.
\newblock {Some quantitative methods for the study of spontaneous activity of
  single neurons}.
\newblock {\em Biophys. J.}, 2:351--368, 1962.

\bibitem{traub}
{Traub RD}, {Whittington MA}, {Colling SB}, {Buzsaki G}, and {Jeffreys JGR}.
\newblock {Analysis of gamma rhythms in the rat hippocampus in vitro and in
  vivo.}
\newblock {\em J Physiol}, 493:471--484, 1996.

\bibitem{Moortgat98}
{Moortgat KT}, {Keller CH}, {Bullock TH}, and {Sejnowski TJ}.
\newblock {Submicrosecond pacemaker precision is behaviorally modulated: the
  gymnotiform electromotor pathway}.
\newblock {\em Proc. Natl. Acad. Sci.}, 95:4684--4689, 1998.

\bibitem{Gluck96}
{Gluckman BJ}, {Netoff TI}, {Neel EJ}, {Ditto WL}, {Spano ML}, and {Schiff SJ}.
\newblock {Stochastic resonance in a neuronal network from a mammalian brain}.
\newblock {\em Phys. Rev. Lett.}, 77:4098--4101, 1996.

\bibitem{Laurent94}
{Laurent G} and {Davidowitz H}.
\newblock {Encoding of Olfactory information with oscillating neural
  assemblies}.
\newblock {\em Science}, 265:1872--1875, 1994.

\bibitem{Laurent96b}
{Wehr M} and {Laurent G}.
\newblock {Odor-encoding by temporal sequences of firing in neural assemblies}.
\newblock {\em Nature}, 384:162--165, 1996.

\bibitem{Laurent98}
{MacLeod K} and {Laurent G}.
\newblock {Who reads temporal information contained across synchronized and
  oscillatory spike trains}.
\newblock {\em Nature}, 395:693--698, 1998.

\bibitem{Wilson98}
{Stern EA}, {Jeager D}, and {Wilson CJ}.
\newblock {Membrane potential of simultaneously recorded striatal spiny neurons
  in vivo}.
\newblock {\em Nature}, 394:475--478, 1998.

\end{thebibliography}

\newpage 
\onecolumn

\newpage

\newpage
\newpage
\newpage
\newpage

\newpage
\begin{figure}
\unitlength=0.1in
\begin{picture}(40,80)
\includegraphics{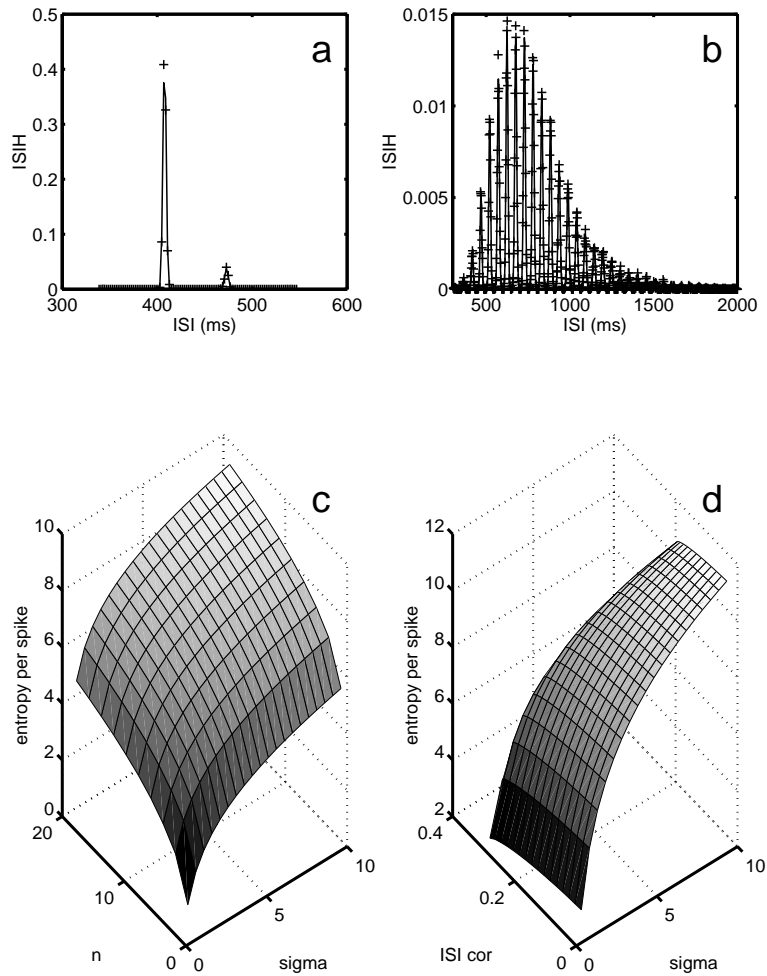}
\end{picture}
\caption{ We plot the ISIH ($+$) and
the corresponding SOG (continuous line) for (a) $D=0.008$ and (b) $D=0.8$,
both for $N=1000$, and $\tau_s=16$.
We show the entropy per spike
(c) as a function of the number of states $n$ and the 
jitter sigma; (d) as a function of the amount of
correlation between consecutive ISI and sigma.
}
\label{Fig7}
\end{figure}
\newpage

\newpage
\appendix

\section{Model equations}

In this appendix we more specifically define the models
studied in this paper.
The dynamics of the voltage $V$ and the kinetic variables $h$, $s$, and $m$ are given by the following
equations,
\begin{eqnarray}
C_m \frac{dV}{dt}&=&-I_{Ca}-I_L-I_{syn}-C_m\xi,\nonumber\\
\frac{dh}{dt}&=&(h_{\infty}-h)/\tau_h,\nonumber\\
\frac{ds}{dt}&=&k_f F(V_{pre})(1-s)-s/\tau_{syn},\nonumber\\
m&=&m_{\infty}(V).\nonumber
\end{eqnarray}
Here the currents are
\begin{eqnarray}  
I_L&=&g_l(V-E_L),\nonumber\\
I_{Ca}&=&g_{Ca}m_{\infty}(V)h(V-E_{Ca}),\nonumber\\
I_{syn}&=&g_{syn}s_{tot}(V-E_{syn}). \nonumber
\end{eqnarray}
The asymptotic values of the kinetic variables, and the $h$ time-scale,
 are specified by
\begin{eqnarray}
m_{\infty}&=&1/[1+\exp -(V+40)/7.4],\nonumber\\
h_{\infty}&=&1/[1+\exp (V+70)/4)],\nonumber\\
\tau_h&=&\phi(\tau_0+\tau_1/[1+\exp (V+50)/3]).\nonumber\\
\end{eqnarray} 
The synaptic activity is implemented in a standard way \cite{wangrinzel93}.
$F$ is chosen such that a presynaptic depolarization higher than $-35~\mbox{mV}$ will open the synaptic channels.
\begin{equation}
F(V)=1/[1+\exp -(V+35)/2].\nonumber
\end{equation} 
The total synaptic coupling in our all-to-all network is defined as
\begin{equation}
s_{tot}=\frac{1}{N}\sum_i s_i,
\end{equation}
and is the same for each neuron.

The standard physiological set of parameters we use in our 
calculations have the conductances
$g_L=0.4$, $g_{Ca}=1.5$, and $g_{syn}=2.0$ (in mS/cm$^2$),
the reversal potentials $E_L=-70$, $E_{Ca}=90$, $E_{syn}=-85$ (in mV),
and $C_m=1 \mu F/cm^2$, $\tau_0=30$, $\tau_1=500$ (in ms),
$k_f=0.5 ms^{-1}$, and $\phi=1.3$. 

\section{Shannon entropy calculation}
As mentioned in the main body of the paper, we quantify the information content in the ISI
using the Shannon entropy,
\begin{equation}
S=-\int {\cal D}(\{ t^n_i\} ) {\cal P}(\{ t^n_i\}) \log_2 {\cal P}(\{ t^n_i\}),
\end{equation}
for the distribution of the spiking times.
In this expression $t^n_i$ is the $i$th spike time of the $n$th neuron,
and ${\cal D}$ is the sum over all spiking time possibilities within a $[0,T]$
interval.
We do not try to evaluate this quantity for the whole network, since at
present it is a very hard calculation to do.
Here we will instead calculate the one neuron entropy in the network, which is
expressed in terms of the distribution $P(\{ \tau_n \})$ of interspike intervals $\tau_n$, as
\begin{eqnarray}
S&=&-\int {\cal D}(\{ \tau_n\} ) p(\{ \tau_n\}) \log_2 p(\{ \tau_n\})\nonumber\\
&\approx& -\langle N \rangle \int~ d\tau P_{ISIH}(\tau)\log_2 P_{ISIH}({\tau}),
\label{ISIHentropy}
\end{eqnarray}
with $\langle N \rangle$ the average number of events during time interval $T$.
We also assume that consecutive ISIs are randomly independent,
\begin{equation}
 p(\tau_1,\cdot\cdot,\tau_N)=\Pi_n P_{ISIH}(\tau_n).
\end{equation}
As explained in the main body of the text the ISIH obtained from our simulations is,
to a good approximation, a sum of Gaussians (SOG), i.e.,
\begin{equation}
P_{ISIH}(\tau)=\sum_i c_i G(\tau|\mu_i,\sigma_i),
\label{SOG}
\end{equation}

where 
\begin{equation} 
G(\tau|\mu_i,\sigma_i)=\frac{1}{\sqrt{2\pi\sigma_i^2}} 
\exp[-(\tau-\mu_i)^2/2\sigma_i^2], 
\end{equation}
with average $\mu_i$,
standard deviation $\sigma_i$, and $c_i$ is the relative weight for
the $i$-th Gaussian contribution.
From the calculated ISIH, we estimate these parameters from
the weight $c_i$, the average $\mu_i$, and the standard deviation $\sigma_i$ of the $i$th peak.
When $(\mu_{i+1}-\mu_i)\gg\sigma_i$, for all $i$, Eq.$~$(\ref{ISIHentropy})
reduces to
\begin{eqnarray}
S&=&\langle N \rangle (\sum_i c_i \log_2 2\pi e \sigma_i^2 -\sum_i c_i
 \log_2 c_i) \nonumber\\
&\equiv& S_1+S_2.
\label{SOGentropy}
\end{eqnarray}

These are the two contributions to the entropy of a multimodal ISIH.
A contribution $S_1$ due to the jitter around the average ISI value of 
a given state $i$, and the contribution $S_2$ due to the discrete probability
distribution of the number of cycles between two spikes.
Note that $S_1$ depends on the
accuracy with which the ISI can be recorded and detected (taken
to be $1~\mbox{ms}$ here). In the other limit $(\mu_{i+1}-\mu_i)<\sigma_i$,
we numerically evaluate the entropy from Eq.$~$(\ref{ISIHentropy}) using the 
measured (binned) ISIH.

In Eq.$~$(\ref{ISIHentropy}) we have assumed
that consecutive ISIs are independent. If consecutive ISIs are correlated, however,
one can use the theory of Markov chains to evaluate the Shannon entropy
of the spike trains. 
The ISIs can belong to $n$ different states, and they have
an equilibrium probability $P_{eq}=(c_1,\cdot\cdot,c_n)$
to be in any particular state. From a given state $i$
the ISI can jump to a new state $j$ with probability $T_{ji}$,
which we can obtain from a return map.
In the return map we plot the next ISI versus the current ISI,
and then divide it into a two-dimensional set of bins $b_{ji}$.
The bins are centered on multiples of the cycle length, and their
width is also equal to the cycle length, then
\begin{equation}
T_{ji}=b_{ji}/\sum_{i} b_{ji},
\end{equation}
and when the ISIs are independent, $T_{ji}=c_j$ (i.e. the
next state does not depend on the previous state).
The actual ISI in each state also displays some jitter,
and it is distributed according to some $\phi_i(\tau)$. We will
assume that $\phi_i$ is Gaussian (as before), and the entropy is then
given by:
\begin{eqnarray}
S&=&-\frac{1}{\log 2} [ (m+1) I\cdot(T\alpha)\cdot P_{eq} 
+\sum_i c_i \log c_i+m  I\cdot(T\log T)\cdot P_{eq} ] \nonumber \\
&\equiv&S_1+S_2+S_3.
\label{MARKOV}
\end{eqnarray}
We have used the following definitions: $m+1$ is the number of ISIs,
$(T\alpha)_{ij}=T_{ij}\alpha_i$ $(T\log T)_{ij}=T_{ij} \log T_{ij}$ (no summations implied), $\alpha_i=\int d\tau \phi_i(\tau) \log \phi_i(\tau)$,
and $\cdot$ denotes matrix multiplication.
Here we give the explicit formula's for $n=2$

\begin{equation}
(T\alpha)\equiv \left(\begin{array}{cc}
\alpha_1 p & \alpha_2 (1-q) \\
\alpha_1 (1-p) & \alpha_2 q \\
\end{array}\right),
\label{Talpha}
\end{equation}

\begin{equation}
(T\log T)\equiv \left(\begin{array}{cc}
 p \log p &  (1-q) \log (1-q)\\
 (1-p) \log (1-p) & q  \log q\\
\end{array}\right).
\label{Logalpha}
\end{equation}

When the ISIs are independent then $S_3=0$, and $S_1$ and $S_2$
reduce to their expression given in Eq.$~$(\ref{SOGentropy}).

We have evaluated expression Eq.$~$(\ref{SOGentropy}) (see Fig.$~$\ref{Fig7}({c})) 
using
\begin{eqnarray}
c_{L+1\pm i}&=&\frac{1}{2\pi}\left(\frac{2}{n}+\sin \frac{2\pi(\pm i +\frac{1}{2})}{n}
-\sin \frac{2\pi(\pm i -\frac{1}{2})}{n}\right),\nonumber\\
\mu_j&=&\mu+(j-L)\Delta\mu,\nonumber\\
\sigma_j&=&\sigma.
\label{CosEnv}
\end{eqnarray}
Here $n=2L+1$, $i=1,\cdot\cdot,L$; $j=1,\cdot\cdot,n$; and $\Delta\mu$ is
the cycle length.

Correlation in consecutive ISIs can be parameterized using a $\delta$,
with $q=1+\delta-p$ in Eqs.$~$(\ref{Talpha}) and (\ref{Logalpha}).
The explicit result for $m=100$ is plotted in Fig.$~$\ref{Fig7}(d).

\end{document}